\documentclass[preprint,11pt]{elsarticle}
\usepackage{graphicx}
\usepackage[linesnumbered,boxed]{algorithm2e}
\usepackage{mathrsfs}
\usepackage{amssymb}
\usepackage{subfigure}
\input amssym.def

\begin{document}

\begin{frontmatter}
\title{Bottom-Left Placement Theorem for Rectangle Packing}
\author{Wenqi Huang, Tao Ye \corref{cor1}}
\cortext[cor1]{Corresponding Author. Tel: 86-27-8754-3885;  Email: yeetao@gmail.com }
\address{School of Computer Science and Technology, Huazhong University of Science and Technology, Wuhan, 430074, China}
\author{Duanbing Chen}
\address{School of Computer Science, University of Electronic Science and Technology of China, Chengdu, 610054, China}
\begin{abstract}
This paper proves a bottom-left placement theorem for the rectangle packing problem, stating that if it is possible to orthogonally place $n$ arbitrarily given rectangles into a rectangular container without overlapping, then we can achieve a feasible packing by successively placing a rectangle onto a bottom-left corner in the container. This theorem shows that even for the real-parameter rectangle packing problem, we can solve it after finite times of bottom-left placement actions. Based on this theorem, we might develop efficient heuristic algorithms for solving the rectangle packing problem.
\end{abstract}

\end{frontmatter}

\section{Introduction}
We consider the following Rectangle Packing (RP) problem: Given a set $J=\{1,2,\cdots,n\}$ of $n$ rectangles, each having width $ w_j$ and height $h_j$, and a rectangular container of width $W$ and height $H$, ask whether it is possible to orthogonally place all rectangles into the container without overlapping. If it is possible, we should answer yes and present a non-overlapping packing pattern; otherwise we should answer no. Note that: (1) $w_j$, $h_j$, $W$ and $H$ are positive \textit{real} numbers. (2) Rectangles are rotatable, i.e., each rectangle can be horizontally or vertically placed into the container. 

The rectangle packing problem arises in many industrial applications, such as cutting wood, glass, paper and  steel in manufacturing, packing goods in transportation, arranging articles and advertisements in publishing. Various algorithms have been proposed to solve this problem. They can be divided into three categories: approximate algorithms, heuristic algorithms and exact algorithms \cite{hopper2001,lodi2002}. 

Most algorithms solve the RP problem by successively placing an unpacked rectangle into the container. Then a basic problem arises:  where to place a new rectangle  when the container is already partially occupied by some packed rectangles?  Generally, there are two approaches  to handle this problem. The first approach  assumes that the parameters of the RP problem are integers and  rectangles can only be placed at some lattice sites \cite{beasley1985,  christofides1977, hadji1995,martello2000}. This approach enumerates all possible positions for a rectangle, thus no feasible solution will be missed.  On the other hand,  the number of possible positions for a rectangle is usually very large and this approach is not applicable to the real-parameter RP problem.  The second approach adopts some placement heuristic which specifies several candidate positions for a rectangle \cite{baker1980, berkey1987, burke2004, huang2011, huang2007}. Using placement heuristic is usually more efficient because the number of candidate positions for a rectangle is much smaller, and it is also applicable to the real-parameter RP problem. However, a risk conceals under the placement heuristic: when there exist feasible solutions, can we achieve one by successively placing a rectangle into the container using the placement heuristic? If the answer is yes, then we say the placement heuristic is \textbf{complete}; otherwise, \textbf{incomplete}. If a placement heuristic is incomplete, then any algorithm based on it is foredoomed to fail on some instances. For example,  \citet{baker1980}  have proved that the Bottom-Left heuristic, which places a rectangle  onto the lowest possible position and left-justify it, is incomplete. They found  an instance for which any feasible solution can not be achieved using the Bottom-Left heuristic, no matter what ordering of the rectangles is used.  \citet{martello2000} proposed a placement heuristic and developed an exact algorithm for the three dimensional bin packing problem. Later, \citet{boef2005} found that  some instances  can not be solved using the placement heuristic proposed by \citet{martello2000}.

It is usually very difficult to prove a placement heuristic's completeness or incompleteness. Nevertheless, for a special case of the RP problem, the 2D rectangular perfect packing problem, \citet{lesh2004} have shown that the  Bottom-Left heuristic is complete. They presented the following theorem: \textit{For every perfect packing, there is a permutation of the rectangles that yields that packing using the Bottom-Left heuristic.}  An efficient branch and bound algorithm is also developed based on this theorem. Besides this result, we have not found other paper in literature proving a certain placement heuristic's completeness. Particularly, no placement heuristic has been rigorously proved to be complete for the general RP problem. 
 
This paper gives a placement heuristic and rigorously proves its completeness. We present the following bottom-left placement theorem:  \textit{ if it is possible to orthogonally place $n$ arbitrarily given rectangles into a rectangular container without overlapping, then we can achieve a feasible packing  by successively placing a rectangle onto a bottom-left corner.} This theorem lays a solid foundation for many efficient and exact algorithms for solving the RP problem\cite{christofides1977,hadji1995, kenmochi2009}.  It is also possible to develop efficient and effective heuristic algorithms based on this theorem.

The rest of the paper is organized as follows. Section 2 presents several notations and definitions. Section 3 proves the bottom-left placement theorem.  Finally, Section 4 concludes this paper and presents some open problems.
\section{Notations and definitions}
We designate the bottom-left corner point of the container as the origin of the $xy$-plane and let its four sides parallel to $x$ and $y$ axis, respectively. A placement of rectangle $i (i=1,2,\cdots, n)$ in the container can be described by three variables $(x_i, y_i, v_i)$, where $x_i, y_i \in \Bbb{R} $ is the coordinate of its bottom-left corner point, $v_i  \in \{0,1\}$ denotes  its orientation, $v_i = 1$ means it is vertically placed, $v_i = 0$ otherwise. A packing pattern of $n$ rectangles can be described by a vector of $3n$ elements: $\mathcal{X}=(x_1, y_1, v_1, x_2, y_2, v_2,\cdots, x_n, y_n, v_n)$. We give the following definitions.

\newdefinition{mydef}{Definition}

\begin{mydef}[Feasible Packing]
A feasible (or non-overlapping) packing must satisfy the following three conditions:
\begin{enumerate}[(1)]
    \item Each rectangle must be orthogonally placed into the container.
    \item Each rectangle must not overstep each border of the container.
    \item The overlapping area between any two rectangles must be zero.
\end{enumerate}
\end{mydef}

\begin{figure}
\begin{center}
\includegraphics[width=1.8in]{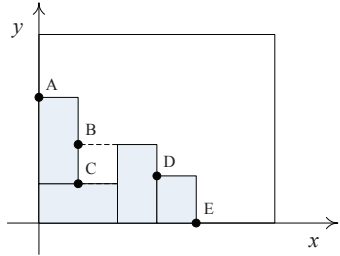}
\caption{Bottom-left stability and bottom-left corners}
\label{fig:blc}
\end{center}
\end{figure}

\begin{mydef}[Bottom-Left Stability]
In a feasible packing, a rectangle is \textit{ bottom-left stable} if and only if it can not move downwards or leftwards without overlapping others \cite{chazelle1983}. A feasible packing is \textit{bottom-left stable} if and only if each rectangle in this packing is bottom-left stable. See Fig.\ref{fig:blc}, each rectangle in the depicted packing has bottom-left stability and  the packing is bottom-left stable.
\end{mydef}

\begin{mydef}[Bottom-Left Corner]
 A bottom-left corner is an unoccupied position where an advisably large rectangle has bottom-left stability. See Fig.\ref{fig:blc}, there are in total five bottom-left corners: A, B, C, D and E.
\end{mydef}

\begin{mydef}[Bottom-Left Placement Action]
A bottom-left placement action is an action that places a rectangle onto a bottom-left corner and makes that rectangle bottom-left stable.
\end{mydef}

\begin{figure}
\begin{center}
\includegraphics[width=1.2in]{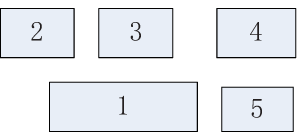}
\caption{Rectangles 2 and 3 are over rectangle 1}
\label{fig:j-over-i}
\end{center}
\end{figure}

\begin{mydef}[Rectangle $j$ Over Rectangle $i$]
We say rectangle $j$ is over rectangle $i$  if and only if there exists a positive real number $d$ such that if rectangle $i$ moves upwards by a distance of $d$,  then the overlapping area between rectangles $i$ and $j$ is greater than zero.  See Fig.\ref{fig:j-over-i}, rectangles 2 and 3 are over rectangle 1, rectangles 4 and 5 are not over rectangle 1. We say rectangle $i$  \textit{ can move upwards freely} if and only if no rectangle is over rectangle $i$. 
\end{mydef}

\begin{figure}
\begin{center}
\includegraphics[height=0.9 in]{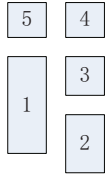}
\caption{Rectangles 2 and 3 are on the right of  rectangle 1}
\label{fig:j-on-the-right-of-i}
\end{center}
\end{figure}

\begin{mydef}[Rectangle $j$ On the Right of Rectangle $i$]
We say rectangle $j$  is on the right of rectangle $i$ if and only if there exists a positive real number $d$ such that if rectangle $i$ moves rightwards by a distance of $d$,  then the overlapping area between rectangles $i$ and $j$ is greater than zero. See Fig.\ref{fig:j-on-the-right-of-i}, rectangles 2 and 3 are on the right of rectangle 1, rectangles 4 and 5 are not on the right of rectangle 1.  We say rectangle $i$ \textit{can move rightwards freely} if and only if no rectangle is on the right of rectangle $i$. 
\end{mydef}

\section{Bottom-left placement theorem}
This section proves the bottom-left placement theorem. We first present two important lemmas.

\newtheorem{mylem}{Lemma}
\newproof{myprf}{Proof}

\begin{mylem}
Any feasible packing can be replaced by another feasible packing where each rectangle has bottom-left stability.
\end{mylem}

\begin{figure}
\begin{center}
 \subfigure[Not bottom-left stable]{
        \includegraphics[width=1.4in]{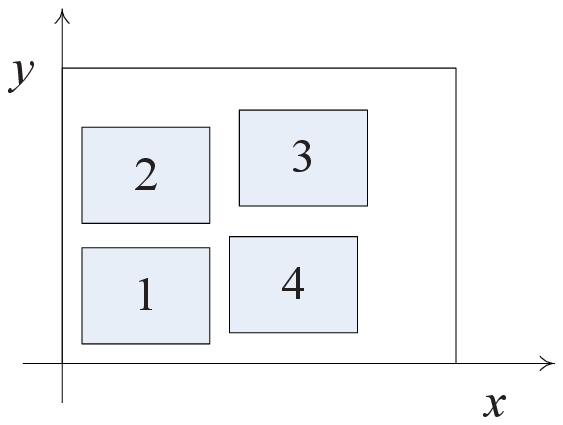}}
  \subfigure[Bottom-left stable]{
        \includegraphics[width=1.4in]{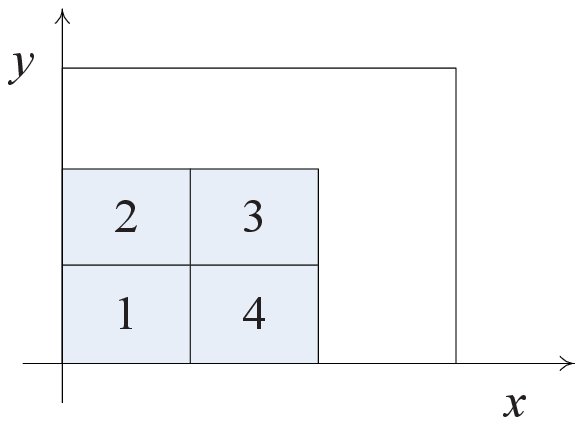}}
  \end{center}
  \caption{A feasible packing and its equivalent bottom-left stable packing}
  \label{fig:lemma1}
\end{figure}

\begin{myprf}
See Fig.\ref{fig:lemma1}, the left packing can be replaced by the right bottom-left stable one. Given a feasible packing $\mathcal{X}_0$, we prove that an equivalent bottom-left stable packing can be found from $\mathcal{X}_0$. Keep the orientation of each rectangle unchanged, suppose that each rectangle can move freely and consider the following function:
\begin{equation}
O = \sum_{i=0}^{n-1}{ \sum_{j=i+1}^{n}{O_{ij}} }
\end{equation}
where $O_{ij} (i,j=1,2,\cdots,n)$ is the  overlapping area between rectangles $i$ and $j$. $O_{0i} (i=1,2,\cdots, n)$  is the overlapping area between rectangle $i$ and the \textbf{outside} of the container.  $O = O(\mathcal{X}) =  O(x_1, y_1, x_2, y_2, \cdots, x_n, y_n) $ is a continuous function defined on $\Bbb{R}^{2n}$.   Let $S_0$ be a set of zero points of $O$: $S_0 = \{\mathcal{X}  |  O(\mathcal{X}) = 0 \}$. Then each point in $S_0$ corresponds to a non-overlapping packing.  $S_0$ is a non-empty, closed and bounded set  because:
\begin{itemize}
  \item  $\mathcal{X}_0$  is  a zero point of $O$, so $S_0$ is not empty.
  \item  $O$ is a continuous function, thus the limit of a sequence of zero points of $O$ is also a zero point, which implies $S_0$ is a closed set. 
  \item In any feasible packing, each rectangle must not overstep each border of the container, so $S_0$ is a bounded set.  
\end{itemize}
Then let's consider a continuous function defined on $S_0$:
\begin{equation}
   L = \sum_{i=1}^{N} (x_i + y_i)
\end{equation}
According to real analysis, a continuous function over a non-empty, closed and bounded set must attain its minimum. Let $\mathcal{X}^* =(x_1^*, y_1^*, x_2^*, y_2^*, \cdots, x_n^*, y_n^*)$ be the point in $S_0$ where $L$ attains its minimum. $\mathcal{X}^*$ corresponds to a feasible packing where each rectangle can not move downwards or leftwards without overlapping others; otherwise, we can find another point in $S_0$ with a smaller $L$, contradicting  the fact that $L$ attains its minimum on $\mathcal{X}^*$.  $\square$
\end{myprf}

Note that Lemma 1 has been explicitly mentioned by \cite{christofides1977, hadji1995, martello2000}  and implicitly used by almost all algorithms for solving the rectangle packing problem. However, to the best of our knowledge, a rigorous and formal proof is first presented here. 

\begin{mylem}[Escaping Lemma]
In any feasible packing, if we take away the four borders of the container, then there is a rectangle which can move upwards and rightwards freely.
\end{mylem}

\begin{figure}
\begin{center}
  \subfigure[]{
      \includegraphics[width=1.4in]{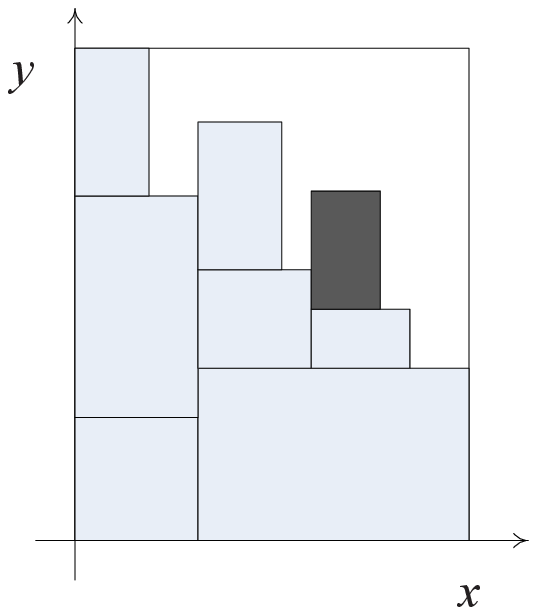}}
  \subfigure[]{
      \includegraphics[width= 1.4in]{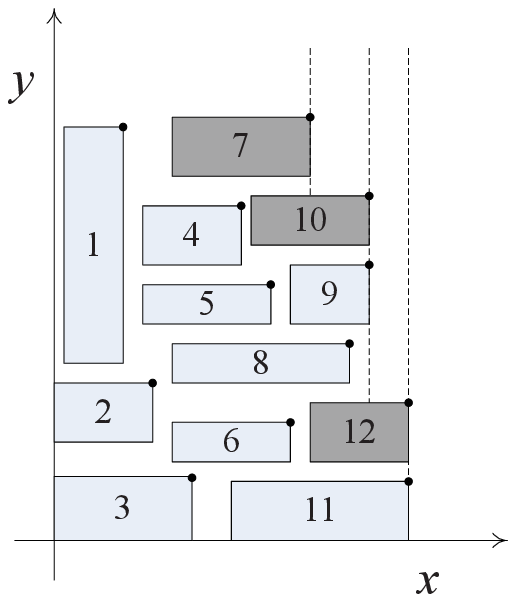}}
  \caption{Examples for Lemma 2}
  \label{fig:lemma2}
\end{center}
\end{figure}
 \begin{myprf}
See Fig.\ref{fig:lemma2}(a), the highlighted rectangle can move  upwards and rightwards freely. Given a feasible packing with $n$ rectangles, we sort the  top-right corner points of the rectangles lexicographically by increasing $<x,y>$ and renumber the rectangles according to this order (See Fig.\ref{fig:lemma2}(b)). We search for the rectangle which can move rightwards and upwards freely as follows. First, we consider the highest numbered rectangle among all $n$ rectangles, i.e., rectangle $n$ (12 in Fig.\ref{fig:lemma2}(b)). Its top-right corner point is the rightmost, thus it can move rightwards freely. If no rectangle is over rectangle $n$,  then $n$ is the rectangle we want to find.  Otherwise,  we consider the highest numbered rectangle among all the rectangles over rectangle $n$. Let it be rectangle $i$ (10 in Fig.\ref{fig:lemma2}(b)). Its top-right corner point is the rightmost among all the rectangles over rectangle $n$. Therefore, it can move rightwards freely. If no rectangle is over rectangle $i$, then $i$ is the rectangle we want to find. Otherwise we consider the highest numbered rectangle among all the rectangles over rectangle $i$ and continue the search as described above.
 
Because there are only $n$ rectangles, the above search will terminate  and we can finally find a rectangle which can move upwards and rightwards freely.  $\square$
\end{myprf}

\newtheorem{mythem}{Theorem}
\begin{mythem}
For any feasible, bottom-left stable packing, there exists a numbering of $n$ rectangles such that rectangle $i(i=1,2,\cdots,n)$ locates on a bottom-left corner formed by rectangles $1,2,\cdots, i-1$ and the four borders of the container.
\end{mythem}

\begin{figure}
\begin{center}
  \subfigure[Numbering According to Lemma 2]{
      \includegraphics[width=1.4in]{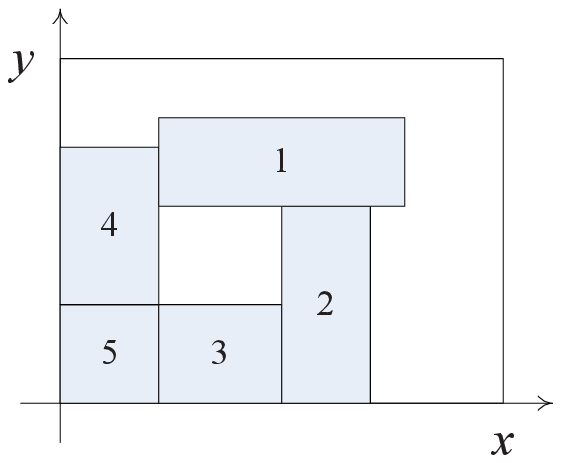}}
  \subfigure[New Numbering]{
      \includegraphics[width= 1.4in]{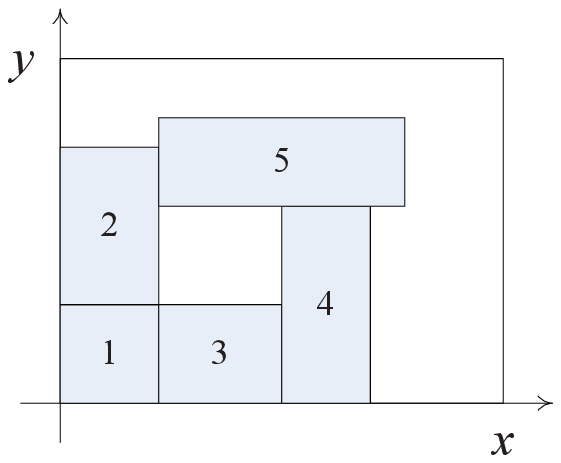}}
  \caption{Examples for Theorem 1}
  \label{fig:theorem1}
\end{center}
\end{figure}

\begin{myprf}
According to Lemma 2, there is always a rectangle which can move rightwards and upwards freely in any feasible packing. Consequently, for a feasible  packing with $n$ rectangles,  we can empty the container by successively taking out a rectangle which can move rightwards and upwards freely. We then number the rectangles according to the order in which rectangles are taken out. Under this numbering, a higher numbered rectangle is not  over or on the right of a lower numbered rectangle (See Fig.\ref{fig:theorem1}(a)).

Now we consider a new numbering which is the reversion of the previous numbering, i.e., the number of rectangle $i(i=1,2,\cdots,n)$ becomes $n-i+1$ (See Fig.\ref{fig:theorem1}(b)). Under this new numbering, a lower numbered rectangle is not over or on the right of a higher numbered rectangle. Then, for  rectangle $i(i=1,2,\cdots,n)$, the labeled numbers of the rectangles forming the bottom-left corner where rectangle $i$ locates are smaller than $i$.

The new numbering of the rectangles can be taken as an order in which rectangles are placed into the container. Under this order, when we place rectangle $i$ into the container, rectangles $1,2,\cdots, i-1$ are already in the container and rectangles $i+1, i+2, \cdots, n$ are not.  Then this theorem shows that, any feasible, bottom-left stable packing can be achieved through a sequence of placement actions, among which the $i$th $(i=1,2,\cdots, n)$ action is to place  rectangle $i$ onto a bottom-left corner formed by rectangles $1,2,\cdots, i-1$ and the four borders of the container. That is to say, any feasible, bottom-left stable packing can be achieved through a sequence of bottom-left placement actions.
$\square$
\end{myprf}

\begin{mythem}[Bottom-Left Placement Theorem]
Arbitrarily given $N$ rectangles and a rectangular container, if it is possible to orthogonally place all rectangles into the container without overlapping, then we can find a feasible packing through a sequence of bottom-left placement actions.
\end{mythem}
\begin{myprf}
According to Lemma 1, there exists a feasible packing where each rectangle has bottom-left stability. Then according to Theorem 1, this bottom-left stable packing can be found through a sequence of bottom-left placement actions. $\square$
\end{myprf}

\begin{mythem}
The rectangle packing problem can be solved after finite times of bottom-left placement actions. That is to say, the rectangle packing problem can be solved after finite times of basic arithmatic and logical operations on the real input parameters $(W,H, w_1, h_1, w_2, h_2,\cdots, w_n, h_n)$.
\end{mythem}
\begin{myprf}
The proofs of Theorems 1 and 2 show that, if it is possible to place all $n$ rectangles into the container without overlapping, then we can find a feasible packing by performing only finite times of bottom-left placement actions. The number of all possible bottom-left placement actions is less than $n!*2^n*(\theta _n)^n$, where $n!$ denotes all possible permutations of $n$ rectangles, $2^n$ all possible orientations of $n$ rectangles, and $\theta _n$ is an upper bound of all possible bottom-left corners for a rectangle.  $n!*2^n*(\theta _n)^n$ is finite, therefore, the search for a feasible packing will finish when all possibilities are checked or once a feasible packing is found.  In the end, if a feasible packing is found, we can answer yes and present the found feasible packing; otherwise, we  can answer no, i.e., it is impossible to orthogonally place all $n$ rectangles into the container without overlapping. 

Note that each bottom-left placement action can be determined by only finite times of basic arithmatic and logical operations, so the theorem is proved. $\square$
\end{myprf}

\section{Open problem}
In this paper, we have proved a bottom-left placement theorem stating that if there exist feasible packings for the RP problem, then we can achieve one by successively placing a rectangle onto a bottom-left corner.  In the future, we want to develop efficient heuristic algorithms for the RP problem based on this theorem. 

Finally, we may present an open problem: Is there an analogue of the bottom-left placement theorem for the 3D rectangle packing problem?


\end{document}